\documentclass[conference,a4paper]{IEEEtran}

\usepackage{multirow}
\usepackage{psfrag}
\usepackage{graphicx}
\usepackage{amsmath}
\newtheorem{theorem}{Theorem}
\newtheorem{lemma}{Lemma}

\begin{document}

\sloppy

\title{Confidence Intervals for the Mutual Information}

\author{
  \IEEEauthorblockN{Arno G. Stefani, Johannes B. Huber}
  \IEEEauthorblockA{Institute for Information Transmission (LIT)\\
    FAU Erlangen-Nuremberg\\
    Erlangen, Germany\\
    Email: \{stefani, huber\}@LNT.de} 
  \and
  \IEEEauthorblockN{Christophe Jardin, Heinrich Sticht}
  \IEEEauthorblockA{Bioinformatics, Institute for Biochemistry\\
    FAU Erlangen-Nuremberg\\
    Erlangen, Germany\\
    Email: \{christophe.jardin, h.sticht\}@biochem.uni-erlangen.de}
}
\maketitle

\begin{abstract}
  "THIS PAPER IS ELIGIBLE FOR THE STUDENT PAPER AWARD"

  By combining a bound on the absolute value of the difference of mutual
  information between two joint probability distributions with a fixed
  variational distance, and a bound on the probability of a maximal
  deviation in variational distance between a true joint probability
  distribution and an empirical joint probability distribution,
  confidence intervals for the mutual information of two random variables
  with finite alphabets are established. Different from previous results,
  these intervals do not need any assumptions on the distribution and the
  sample size.
\end{abstract}

\section{Introduction}

In this paper confidence intervals for the mutual information of two
random variables with finite alphabets are established. While they
are not particularly tight, they are the first where no further restrictions
have to be considered, neither on being in an asymptotic regime nor on the underlying joint
probability distribution. By quantization of random variables with a non
finite alphabet it is also possible to find the lower bound of the
confidence interval of the mutual information of such random variables. The
simplicity of these confidence intervals also allows to give an upper
bound on the necessary sample size when the confidence interval width, the confidence
level, and the alphabet sizes are fixed.

\section{Notational Setup}

Let $X$, $Y$, $X'$, $Y'$ be two pairs of finite discrete random variables,
with joint probability distributions 
\begin{align}
p_{XY} &{}={} \{ p_{XY}(i,j) : i=1, 2, \ldots, M_x;~j=1, 2, \ldots, M_y \},\nonumber\\
p_{X'Y'} &{}={} \{ p_{X'Y'}(i,j) : i=1, 2, \ldots, M_x;~j=1, 2, \ldots, M_y \}. \nonumber
\end{align}
Here $X, X' \in \mathcal{X}$ and $Y, Y' \in \mathcal{Y}$ and it is w.l.o.g.
assumed that $\mathcal{X}=\{ 1, 2, \ldots, M_x \}$ and that $\mathcal{Y}=\{ 1, 2, \ldots, M_y \}$.
The marginal probability distributions are
$p_X = \{ p_X(i) : i=1, 2, \ldots, M_x \}$, $p_Y = \{ p_Y(j) : j=1, 2, \ldots,
M_y \}$, $p_{X'} = \{ p_{X'}(i) : i=1, 2, \ldots, M_x \}$ and $p_{X'} = \{
p_{Y'}(j) : j=1, 2, \ldots, M_y \}$, where the marginals are calculated from
the joint probalility distributions as usual.
The Shannon entropy \cite{Cover2006} is defined as 
\[
H(X) = H(p_X) = -\sum\limits_{i=1}^{M_x} p_X(i) \log p_X(i)
\]
and the joint entropy \cite{Cover2006} as
\[
H(XY) = H(p_{XY}) = -\sum\limits_{i=1}^{M_x}\sum\limits_{j=1}^{M_y} p_{XY}(i,j) \log p_{XY}(i,j).
\]
All $\log$s are natural if not stated otherwise. $\mathcal{H}(\cdot)$ is defined as the
binary entropy function
\[
\mathcal{H}(x)=-x \log x -(1-x) \log(1-x).
\]
The mutual information \cite{Cover2006} is defined as
\begin{align}
\label{eqn_MI}
I(X;Y) = I(p_{XY}) = H(X) + H(Y) - H(XY).
\end{align}
W.l.o.g. it is assumed, that $M_x \leq M_y$, what can be done because the mutual
information is symmetric ($I(X;Y)=I(Y;X)$), and therefore by renaming the
variables if necessary it can be assumed that $M_x \leq M_y$ always holds.
The variational distance between two probability distributions is defined as
\begin{align}
V(p_{XY}, p_{X'Y'}) &= \left\Vert p_{XY} - p_{X'Y'} \right\Vert_1 \nonumber\\
		    &= \sum\limits_{i=1}^{M_x}\sum\limits_{j=1}^{M_y} | p_{XY}(i,j) - p_{X'Y'}(i,j) |, \nonumber
\end{align}
and similarly for the marginal distributions. It can be easily seen, that
$V(\cdot,\cdot) \in [0,2]$ for any two probability distributions.
The empirical joint distribution for an i.i.d. sequence of pairs $((x_1, y_1), (x_2, y_2), \ldots,
(x_n, y_n))$, sampled from a distribution $p_{XY}$, is defined as 
\begin{align}
p_{x^ny^n} = \{ p_{x^ny^n}(i,j) : i=1, 2, \ldots, M_x;~j=1, 2, \ldots, M_y \},\nonumber
\end{align}
where
\begin{align}
p_{x^ny^n}(i,j) = \frac{1}{n} \sum\limits_{k=1}^{n} \delta_{x_k i} \delta_{y_k j}
\end{align} 
and $\delta_{ij}$ is the Kronecker delta.

\section{Related Work}

The following two bounds will be used to construct the confidence interval
for mutual information and are stated here as two Lemmas. 
\begin{lemma}
\label{lemma1}
Let $(X, Y)$ and $(X', Y')$ be two pairs of random variables taking values
on the same range, with joint probability distributions $p_{XY}$ and $p_{X'Y'}$.
Let
\[
\epsilon = V(p_{XY},p_{X'Y'}).
\]
If $\epsilon \leq 2-\frac{2}{M_x M_y}$, then it holds that
\begin{align}
| I(X;Y) - I(X'&;Y') | \nonumber\\
	 & \leq 3 \cdot \frac{\epsilon}{2} \log(M_x M_y-1) + 3 \mathcal{H}(\tfrac{\epsilon}{2}). \label{eqn_Zhang}
\end{align}
\end{lemma}
\begin{lemma}
\label{lemma2}
For any $\epsilon > 0$
\begin{align}
\label{eqn_Weissman}
\mathrm{Pr}\{V(p_{XY},p_{X^nY^n}) > \epsilon\} < (2^{M_xM_y}-2)\mathrm{e}^{-n \epsilon^2/2}.
\end{align}
\end{lemma}

The first
bound was found by Zhang \cite[Theorem 2]{Zhang2007}. In the next section this
bound will be slightly improved and generalized for the usage here, using
a result of Ho and Yeung \cite[Theorem 6]{Ho2010}. The second bound was
originally found by Weissman et al. \cite[Theorem 2.1]{Weissman2003} and
slightly modified by Ho and Yeung \cite[Lemma 3]{Ho2010} to have no dependence
on the true distribution.

\section{Results}
\label{sec_results}

First, (\ref{eqn_Zhang}) is improved to yield:
\begin{theorem}
\label{Zhang_improved}
Let $(X, Y)$ and $(X', Y')$ be two pairs of random variables taking values
on the same range, with joint probability distributions $p_{XY}$ and $p_{X'Y'}$
and $M_x \leq M_y$. Fix an $\epsilon > 0$.
Let
\[
V(p_{XY},p_{X'Y'}) \leq \epsilon.
\]
Then it holds that
\begin{align}
| I(X&;Y) - I(X';Y') | \nonumber\\
	 \label{eqn_Zhang_improved}
	 & \leq \begin{cases} 
		\frac{\epsilon}{2} \log[(M_x M_y-1)(M_x-1)(M_y-1)] + 3 \mathcal{H}(\tfrac{\epsilon}{2}) \\
		~~~~ \text{for }\epsilon \leq 2-\frac{2}{M_x} \\
		\log(M_x) \\
		~~~~ \text{for }\epsilon > 2-\frac{2}{M_x}.
		\end{cases}
\end{align}
\end{theorem}
\begin{IEEEproof}
The proof widely follows the lines of the proof of (\ref{eqn_Zhang}) in
Zhang \cite[Eq. (2)]{Zhang2007}, but replaces the entropy difference bound of
Zhang \cite[Eq. 4]{Zhang2007} by the corresponding bound in Ho and Yeung
\cite[Theorem 6]{Ho2010}, what makes the new bound valid for any $\epsilon$
and also for any $V(p_{XY},p_{X'Y'}) \leq \epsilon$ instead of $V(p_{XY},p_{X'Y'})
= \epsilon$. Beyond this, some slight changes in the proof of Zhang lead to a
tighter bound.

First it is shown that $V(p_{X},p_{X'}) \leq \epsilon:$
\begin{align}
V(p_{X},p_{X'}) &= \left\Vert p_{X} - p_{X'} \right\Vert_1 \nonumber\\
		&= \sum\limits_{i=1}^{M_x} | p_{X}(i) - p_{X'}(i) | \nonumber\\
		&= \sum\limits_{i=1}^{M_x} \left| \sum\limits_{j=1}^{M_y}(p_{XY}(i,j) - p_{X'Y'}(i,j)) \right| \nonumber\\
		&\leq \sum\limits_{i=1}^{M_x} \sum\limits_{j=1}^{M_y} | p_{XY}(i,j) - p_{X'Y'}(i,j) | \nonumber\\
		&= V(p_{XY},p_{X'Y'}) \nonumber\\
		&\leq \epsilon \nonumber
\end{align}
In an analogous way it can be shown that $V(p_Y,p_{Y'}) \leq \epsilon$.

For $\epsilon \leq 2-\frac{2}{M_x}$ then it holds:
\begin{align}
| I(X&;Y) - I(X';Y') | \nonumber\\
     \label{eqn_MI_app}
     &= | H(X) + H(Y) - H(XY) \nonumber\\
     & ~~~~ - H(X') - H(Y') + H(X'Y') | \\
     &\leq | H(X) - H(X') | + | H(Y) - H(Y') | \nonumber\\
     & ~~~~ + | H(XY) - H(X'Y') | \nonumber\\
     \label{eqn_Ho_Th6}
     &\leq \frac{\epsilon}{2} \log(M_x-1) + \mathcal{H}(\tfrac{\epsilon}{2}) + \frac{\epsilon}{2} \log(M_y-1) + \mathcal{H}(\tfrac{\epsilon}{2}) \nonumber\\
     & ~~~~ + \frac{\epsilon}{2} \log(M_xM_y-1) + \mathcal{H}(\tfrac{\epsilon}{2}) \\
     &= \frac{\epsilon}{2} \log[(M_x M_y-1)(M_x-1)(M_y-1)] + 3 \mathcal{H}(\tfrac{\epsilon}{2}) \nonumber
\end{align}
In (\ref{eqn_MI_app}) eq. (\ref{eqn_MI}) was used. In 
(\ref{eqn_Ho_Th6}) the bound of Ho and Yeung \cite[Theorem 6]{Ho2010}
was applied together with the assumption $M_x \leq M_y$ and therefore,
by the assumption $\epsilon \leq 2-\frac{2}{M_x}$, with $2-\frac{2}{M_xM_y}
\geq 2-\frac{2}{M_y} \geq 2-\frac{2}{M_x} \geq \epsilon$.

For $\epsilon > 2-\frac{2}{M_x}$ the well known bounds on mutual information
and entropy \cite{Cover2006}, $I(X;Y) \geq 0$ and $I(X;Y) \leq H(X) \leq \log M_x$ 
are first used to show that
\begin{align}
&0 \leq I(X;Y), I(X';Y') \leq \log M_x, \nonumber
\end{align}
what immediately implies
\begin{align}
| I(X;Y) - I(X';Y') | \leq \log M_x,
\end{align}
independent of $\epsilon$, what completes the proof.
\end{IEEEproof}

\emph{Remark:} The absolute entropy difference bound of Ho and Yeung \cite[Theorem 6]{Ho2010} could also 
be used to bound $| I(X;Y) - I(X';Y') |$ in the case $\epsilon > 2-\frac{2}{M_x}$,
but here it can easily be seen that $| I(X;Y) - I(X';Y') | = | H(X) - H(X') |
+ | H(Y) - H(Y') | + | H(XY) - H(X'Y') | \leq \log M_x + | H(Y) - H(Y') | +
| H(XY) - H(X'Y') | \geq \log M_x$ and therefore the upper bound $\log M_x$ is
tighter for $\epsilon > 2-\frac{2}{M_x}$. From this argumentation it can also be
seen that the upper bound for the case that $\epsilon$ is smaller, but close to $2-\frac{2}{M_x}$,
is still greater than $\log M_x$, and could therefore be improved by
taking the minimum of this bound and $\log M_x$, but for the sake of simplicity and
applicability of this bound this improvement has not been applied in Theorem
\ref{Zhang_improved}. This shows that this bound is only useful for sufficiently
small $\epsilon$, since $\log M_x$ is a well known and in the context of confidence
intervals trivial bound. Nevertheless (\ref{eqn_Zhang_improved}) is everywhere tighter than (\ref{eqn_Zhang}), 
applicable for any $\epsilon$, and the variational distance $V(p_{XY},
p_{X'Y'})$ has only to be \emph{less or equal} $\epsilon$ and not strictly \emph{equal} to $\epsilon$ for
(\ref{eqn_Zhang_improved}). Therefore Theorem 1 is an
improvement of the bound of Zhang (Lemma~\ref{lemma1}).

Finally the confidence interval is constructed by a combination of Theorem
\ref{Zhang_improved} and Lemma \ref{lemma2}.
\begin{theorem}
\label{theorem2}
For any $\alpha \in (0,1]$ and $M_x$, $M_y$ with $M_x \leq M_y$ let (where $\ln$ is the natural logarithm)
\[
\epsilon = \sqrt{\frac{2}{n} \ln\frac{2^{M_xM_y}-2}{\alpha}}
\]
and
\[
\Delta I(\epsilon) = \begin{cases} 
		     \frac{\epsilon}{2} \log[(M_x M_y-1)(M_x-1)(M_y-1)] + 3 \mathcal{H}(\tfrac{\epsilon}{2}) \\
		     ~~~~ \text{for }\epsilon \leq 2-\frac{2}{M_x} \\
		     \log(M_x) \\
		     ~~~~ \text{for }\epsilon > 2-\frac{2}{M_x}
		     \end{cases}
\]
then, for any two random variables $X$, $Y$ with true joint probability
distribution $p_{XY}$ and empirical joint probability distribution $p_{X^nY^n}$
it holds that
\begin{align*}
\mathrm{Pr}\{I(p_{X^nY^n})-\Delta I(\epsilon) \leq I(p_{XY}) \leq I(p_{X^nY^n})+ & \Delta I(\epsilon)\} \\
										& \geq 1-\alpha .
\end{align*}
\end{theorem}
\begin{IEEEproof}
Rewriting (\ref{eqn_Weissman}) as
\begin{align}
\label{eqn_CI_proof1}
\mathrm{Pr}\{V(p_{XY},p_{X^nY^n}) \leq \epsilon\} \geq 1 - (2^{M_xM_y}-2)\mathrm{e}^{-n \epsilon^2/2},
\end{align}
and solving $1-\alpha=1 - (2^{M_xM_y}-2)\mathrm{e}^{-n \epsilon^2/2}$ yields (obviously
only the positive solution is of interest)
\[
\epsilon = \sqrt{\frac{2}{n} \ln\frac{2^{M_xM_y}-2}{\alpha}}.
\]
Then it follows that
\begin{align}
&  1-\alpha \nonumber\\
&\leq \mathrm{Pr}\{V(p_{XY},p_{X^nY^n}) \leq \epsilon\} \nonumber\\
\label{eqn_CI_proof2}
&\leq \mathrm{Pr}\{ | I(p_{X^nY^n})-I(p_{XY}) | \leq \Delta I(\epsilon) \} \\
&= \mathrm{Pr}\{I(p_{X^nY^n})-\Delta I(\epsilon) \leq I(p_{XY}) \leq I(p_{X^nY^n})+ \Delta I(\epsilon)\}, \nonumber
\end{align}
where (\ref{eqn_CI_proof2}) is an application of Theorem \ref{Zhang_improved}.
\end{IEEEproof}

The next theorem gives an upper bound on the necessary number of
samples $n$, to achieve a given confidence interval width at a given
confidence level $1-\alpha$.
\begin{theorem}
\label{theorem3}
For any $\alpha \in (0, 1]$, $M_x$, $M_y$, with $M_x \leq M_y$, and $\gamma
\in (0, \log M_x)$ let $\epsilon$ be the minimum root of
\begin{align}
\label{th3_eqn1}
\frac{\epsilon}{2} \log[(M_x M_y-1)(M_x-1)(M_y-1)] + 3 \mathcal{H}(\tfrac{\epsilon}{2})=\gamma.
\end{align}
Then for ($\lceil \cdot \rceil$ is the ceiling operator)
\[
n = \left\lceil \frac{2}{\epsilon^2} \ln \frac{2^{M_xM_y}-2}{\alpha} \right\rceil
\]
it holds that
\[
\mathrm{Pr}\{I(p_{X^nY^n})-\gamma \leq I(p_{XY}) \leq I(p_{X^nY^n})+ \gamma\} \geq 1-\alpha.
\]
\end{theorem}
\begin{IEEEproof}
If $\gamma \geq \log M_x$ then the probability of being within the bounds is trivially
one, therefore $\gamma$ is restricted to be less $\log M_x$. Then obviously only the
first part of (\ref{eqn_Zhang_improved})
\[
\frac{\epsilon}{2} \log[(M_x M_y-1)(M_x-1)(M_y-1)] + 3 \mathcal{H}(\tfrac{\epsilon}{2})
\]
applies, where $\epsilon \leq 2-\frac{2}{M_x}$. It is easy to show,
that this term is strictly increasing for $\epsilon \in (0, 2-\frac{2}{M_x})$. Therefore
there is only one solution for $\epsilon \in (0, 2-\frac{2}{M_x})$ of equation (\ref{th3_eqn1})
which is just the desired maximal variational distance between the true and the
empirical joint distribution. This $\epsilon$ is also the minimum root as stated in
the theorem. Then solving (\ref{eqn_CI_proof1}) for $n$, after the substitution of
$\mathrm{Pr}\{V(p_{XY},p_{X^nY^n}) \leq \epsilon\}$ by $1-\alpha$, yields
\[
n \geq \frac{2}{\epsilon^2} \ln \frac{2^{M_xM_y}-2}{\alpha}
\]
and therefore
\[
n = \left\lceil \frac{2}{\epsilon^2} \ln \frac{2^{M_xM_y}-2}{\alpha} \right\rceil
\]
cleary suffices to guarantee
\[
\mathrm{Pr}\{I(p_{X^nY^n})-\gamma \leq I(p_{XY}) \leq I(p_{X^nY^n})+ \gamma\} \geq 1-\alpha.
\]
\end{IEEEproof}

The next theorem is an improvement of Theorem \ref{theorem2}, that uses the entropy
optimization procedures of \cite[Theorems 2 and 3]{Ho2010}, which depend on the
actual empirical distribution, instead of the worst case entropy difference bound
\cite[Theorem 6]{Ho2010}.
\begin{theorem}
\label{theorem4}
For any $\alpha \in (0,1]$ and $M_x$, $M_y$ with $M_x \leq M_y$ let
\[
\epsilon = \sqrt{\frac{2}{n} \ln\frac{2^{M_xM_y}-2}{\alpha}}
\]
and let
\begin{align*}
I_\mathrm{min} = & \min\limits_{p_X:~V(p_{X^n},p_X) \leq \epsilon} H(X) + \min\limits_{p_Y:~V(p_{Y^n},p_Y) \leq \epsilon} H(Y) \\
		 & - \max\limits_{p_{XY}:~V(p_{X^nY^n},p_{XY}) \leq \epsilon} H(XY), \\
I_\mathrm{max} = & \max\limits_{p_X:~V(p_{X^n},p_X) \leq \epsilon} H(X) + \max\limits_{p_Y:~V(p_{Y^n},p_Y) \leq \epsilon} H(Y) \\
		 & - \min\limits_{p_{XY}:~V(p_{X^nY^n},p_{XY}) \leq \epsilon} H(XY)
\end{align*}
where the solutions for the entropy optimization problems are given in \cite[Theorems 2 and 3]{Ho2010}. Then it holds that
\begin{align*}
\mathrm{Pr}\{I_\mathrm{min} \leq I(p_{XY}) \leq I_\mathrm{max}\} \geq 1-\alpha.
\end{align*}
\end{theorem}
\begin{IEEEproof}
Since $V(p_{X^n},p_X)$ as well as $V(p_{Y^n},p_Y)$ are $\leq V(p_{X^nY^n},p_{XY}) \leq \epsilon$,
as shown in the proof of Theorem \ref{Zhang_improved}, it is obvious that
\begin{align*}
\min\limits_{p_{XY}:~V(p_{X^nY^n},p_{XY}) \leq \epsilon} I(p_{XY}) & \geq I_\mathrm{min}, \\
\max\limits_{p_{XY}:~V(p_{X^nY^n},p_{XY}) \leq \epsilon} I(p_{XY}) & \leq I_\mathrm{max}.
\end{align*}
By the argumentation of the proof of Theorem \ref{theorem2} again
\[
\epsilon = \sqrt{\frac{2}{n} \ln\frac{2^{M_xM_y}-2}{\alpha}}
\]
is fixed, and it follows that
\begin{align}
&  1-\alpha \nonumber\\
&\leq \mathrm{Pr}\{V(p_{XY},p_{X^nY^n}) \leq \epsilon\} \nonumber\\
&\leq \mathrm{Pr}\{I_\mathrm{min} \leq I(p_{XY}) \leq I_\mathrm{max}\}. \nonumber
\end{align}
\end{IEEEproof}

\section{Discussion}
\label{sec_discussion}

Theorem \ref{theorem3} can be seen as an upper bound for $n$ (the number of samples), which
is tight when Theorem \ref{theorem2} is used to determine the confidence interval. This
is explained by the fact, that the absolute entropy difference bound that was used to construct the confidence intervals is
completely independent of the actual empirical distribution $p_{x^ny^n}$. Also, by using the entropy
difference bounds, the dependence between the entropies $H(X)$, $H(Y)$ and $H(XY)$ was ignored, since for example
the worst case distribution $p_{x^n}$ is not necessarily the marginal of the worst case distribution $p_{x^ny^n}$,
what makes the mutual information difference bound less tight again.

Taken together, one can see that there is much room left for improvement. By this,
$n$ of Theorem \ref{theorem3} is an upper bound on the necessary smaples size.

A first improvement of this situation was given in Theorem~\ref{theorem4}.

An approach for making also use of the dependence between the entropies is given as
a conjeture and only for two binary random variables in \cite{Stefani2012}.

Besides this in the preprint~\cite{Stefani2013}, an algorithm for finding the lower
bound of the confidence interval for a binary and an arbitrary finite 
random variable is given. This bound is tight in terms of the
maximal variational distance between the empirical and the true joint distribution.

\section{Numerical Examples}

In this section the different possibilities for the construction of the
confidence intervals, which just have been discussed are compared in two numerical
examples. In these particular examples it can be seen that the lower bound conjectured
in \cite{Stefani2012} (called Method 1) matches the lower bound of preprint~\cite{Stefani2013} (called Method 2) which gives a further
indication for the correctness of at least the lower bound in \cite{Stefani2012}
(though there is still no proof available). 

\begin{figure}[htbp]
  \centering
  \psfrag{MI}[c][c]{$I(p_{X^nY^n})$}
  \psfrag{cdf}[c][c]{empirical sampling cdf}
  \includegraphics[width=0.5\textwidth]{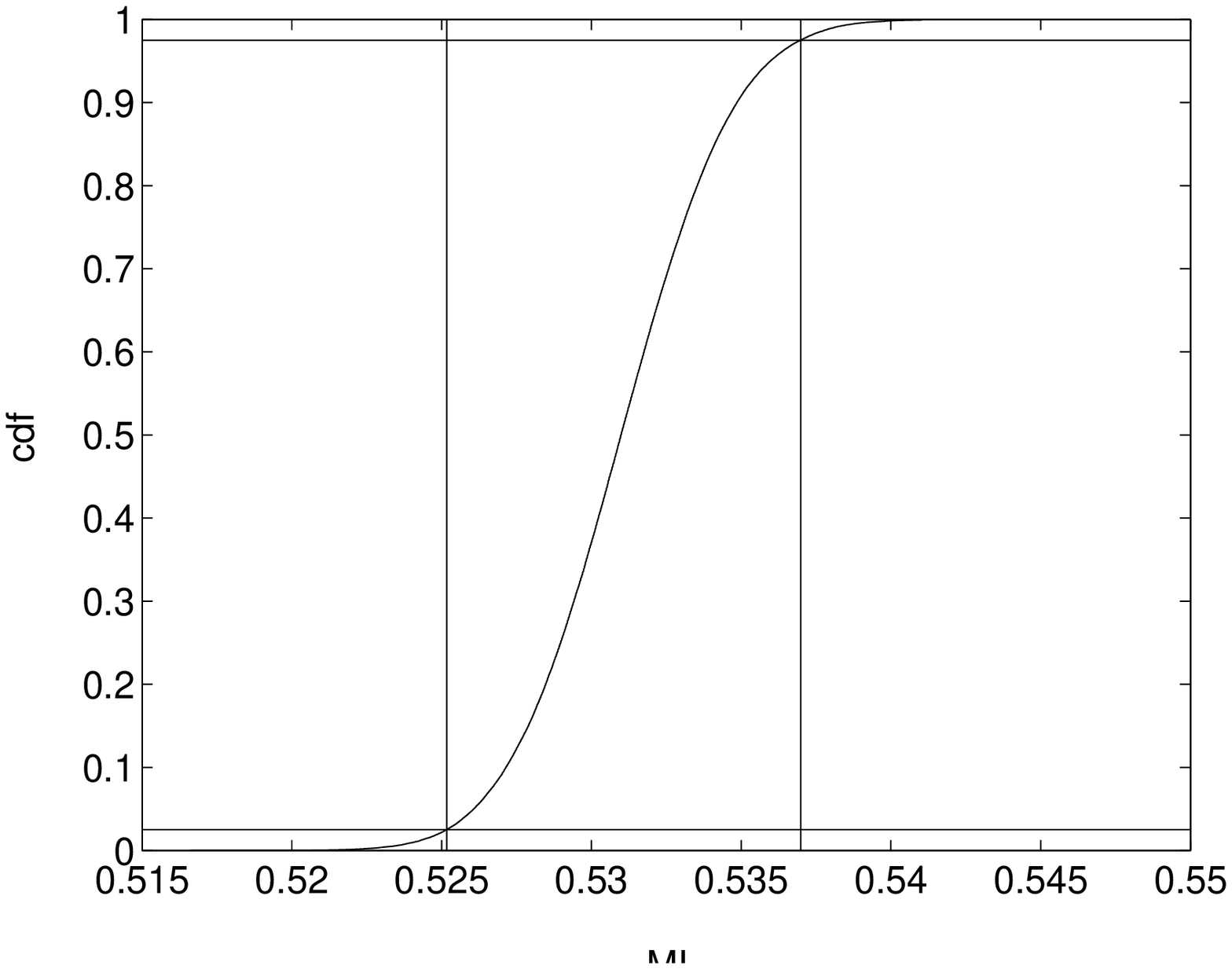}
  \caption{}
  \label{fig:MI_cdf}
\end{figure}

The following setup is used: A binary symmetric channel (BSC) with input variable
$X$ and output variable $Y$ is given, where the bit error rate ($\mathrm{BER}$) is equal to
$0.1$ and the input probabilities $p_X=\{\frac{1}{2},\frac{1}{2}\}$.
\centerline{
\includegraphics{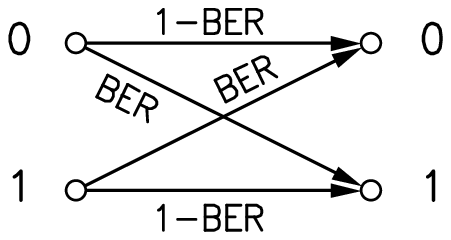}}
The joint probabilities therefore are
\begin{align*}
p_{XY}(1,1)=0.45,~~ & p_{XY}(1,2)=0.05, \\
p_{XY}(2,1)=0.05,~~ & p_{XY}(2,2)=0.45.
\end{align*}
In this case the true mutual information is known to be
\[
I(p_{XY})=1-\mathcal{H}(0.1) \approx 0.53100
\]
(unlike in the sections before, in this section all $\log$s are to the base 2).
Then, taking $n=10^5$ samples from $p_{XY}$ yielded the following exemplary empirical distribution
\begin{align*}
p_{x^ny^n}(1,1)=0.44950,~~ & p_{x^ny^n}(1,2)=0.05058, \\
p_{x^ny^n}(2,1)=0.04868,~~ & p_{x^ny^n}(2,2)=0.45124.
\end{align*}
Now fixing the confidence level $1-\alpha=0.95$ the predescribed methods could be
used to estimate the confidence interval.
Before this is done, a good approximation to the best possible confidence interval is determined, where best possible interval is defined as having minimal interval width. Therefore
samples of size $n$ are sampled $10^5$ times from $p_{XY}$, yielding an exemplary empirical sampling cumulative distribution function (cdf) of $I(p_{X^nY^n})$ (shown in Fig. \ref{fig:MI_cdf}), which should be a sufficiently good approximation to the real sampling cdf of $I(p_{X^nY^n})$, due to the high number of samples.

%

Then, since it can be seen from the empirical sampling cdf of $I(p_{X^nY^n})$ that the sampling probability density function (pdf) is close to being
unimodal and symmetric, the approximation to the smallest possible confidence interval is given by the
$\frac{\alpha}{2}\text{-quantile} \approx  0.52517$ and the $(1-\frac{\alpha}{2})\text{-quantile} 
\approx 0.53699$ of the empirical sampling cdf of $I(p_{X^nY^n})$ (both marked in Fig.~\ref{fig:MI_cdf}).

In Table~\ref{table1} the results of the two methods described in Section~\ref{sec_results}
(Theorem \ref{theorem2} and \ref{theorem4}) and of Method 1 and 2, applied to
$p_{x^ny^n}$, are given.

%
\begin{table}[htbp]
  \renewcommand{\arraystretch}{1.3}
  \caption{}
  \label{table1}
  \centering
  \begin{tabular}{|c||c|c|c|}
    \hline
    \multirow{2}{*}{Method} & \multicolumn{3}{|c|}{Confidence interval} \\
    \cline{2-4}
	   & Lower bound & Upper bound & Width \\
    \hline\hline
    approximated best possible  & 0.52517 & 0.53699 & 0.01182 \\
    \hline
    Theorem 2	   & 0.38170 & 0.68504 & 0.30334 \\
    \hline
    Theorem 4	   & 0.51645 & 0.55091 & 0.03445 \\
    \hline
    Method 1	   & 0.51666 & 0.55080 & 0.03414 \\
    \hline
    Method 2	   & 0.51666 &   ---   &   ---   \\
    \hline
  \end{tabular}
\end{table}

\begin{figure}[htbp]
  \centering
  \psfrag{MI}[c][c]{$I(p_{X^nY^n})$}
  \psfrag{cdf}[c][c]{empirical sampling cdf}
  \includegraphics[width=0.5\textwidth]{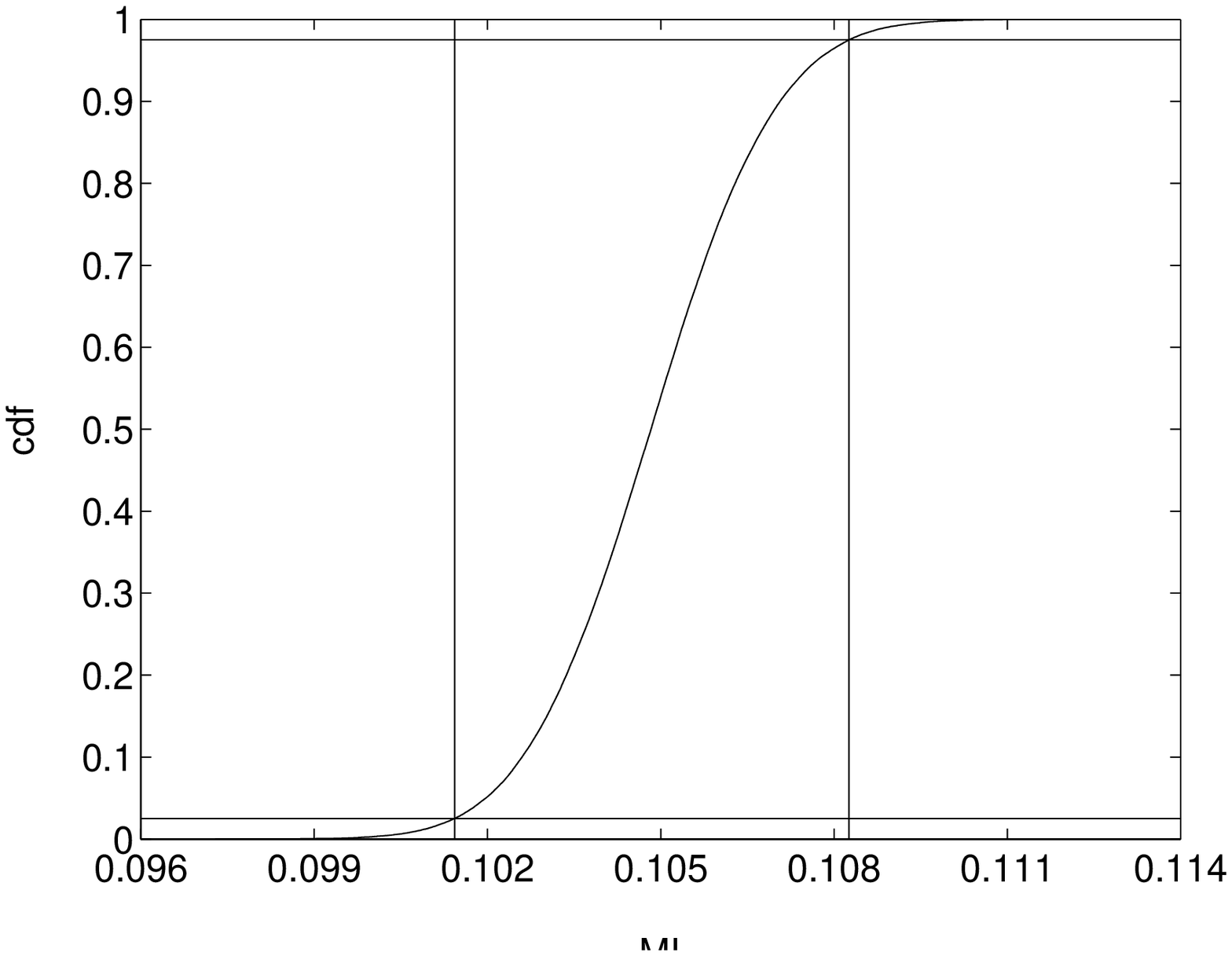}
  \caption{}
  \label{fig:small_MI_cdf}
\end{figure}

Here it can be seen, that the independence of the empirical distribution
in Theorem \ref{theorem2} makes the confidence interval pretty broad compared
to the other methods. Besides this, one can see that the improved methods
(Method 1 and 2 in Table~\ref{table1}) have nearly the same performance as
Theorem \ref{theorem4}. The situation rather changes when a true distribution
with small mutual information is used (such a situation is prevalent in \cite{Othersen2012}).
This is shown in the following example, where a BSC is used with $\mathrm{BER}=0.2$
and an unequally distributed input variable $X$ with distribution $p_X=\{ 0.1, 0.9 \}$.
The joint probabilities therefore are
\begin{align*}
p_{XY}(1,1)=0.08,~~ & p_{XY}(1,2)=0.02, \\
p_{XY}(2,1)=0.18,~~ & p_{XY}(2,2)=0.72.
\end{align*}
Here the true mutual information
\[
I(p_{XY})\approx 0.10482.
\]
Again taking $n=10^5$ samples from $p_{XY}$ yielded the following exemplary empirical joint distribution
\begin{align*}
p_{x^ny^n}(1,1)=0.07996,~~ & p_{x^ny^n}(1,2)=0.02023, \\
p_{x^ny^n}(2,1)=0.18012,~~ & p_{x^ny^n}(2,2)=0.71969.
\end{align*}
The sampling cdf of $I(p_{X^nY^n})$ in this case can be seen in Fig.~\ref{fig:small_MI_cdf}.
The approximation to the smallest possible confidence interval is determined by the same method as in the first example.
The results are given in Table~\ref{table2}.
\begin{table}[htbp]
  \renewcommand{\arraystretch}{1.3}
  \caption{}
  \label{table2}
  \centering
  \begin{tabular}{|c||c|c|c|}
    \hline
    \multirow{2}{*}{Method} & \multicolumn{3}{|c|}{Confidence interval} \\
    \cline{2-4}
	   & Lower bound & Upper bound & Width \\
    \hline\hline
    approximated best possible  &  0.10143 & 0.10826 & 0.00683 \\
    \hline
    Theorem 2	   & -0.04743 & 0.25591 & 0.30334 \\
    \hline
    Theorem 4	   &  0.05269 & 0.15721 & 0.10452 \\
    \hline
    Method 1	   &  0.08679 & 0.12402 & 0.03723 \\
    \hline
    Method 2	   &  0.08679 &   ---   &   ---   \\
    \hline
  \end{tabular}
\end{table}

\section*{Acknowledgment}

The authors would like to thank the DFG for supporting 
their research with SPP1395 in the projects HU634\_7 and STI155\_3.


\begin{thebibliography}{1}
\bibitem{Cover2006}
  T.~M. Cover and J.~A. Thomas, \emph{Elements of Information Theory},
  2nd ed. New York: Wiley, 2006.
\bibitem{Zhang2007}
  Z. Zhang, ``Estimating mutual information via Kolmogorov distance,''
  \emph{IEEE Trans. Inform. Theory}, vol.~53, no. 9, pp. 3280--3282,
  Sep. 2007.
\bibitem{Ho2010}
  S.-W. Ho and R.~W. Yeung, ``The interplay between entropy and 
  variational distance,'' \emph{IEEE Trans. Inform. Theory}, 
  vol.~56, no. 12, pp. 5906--5929, Dec. 2010.
\bibitem{Stefani2012}
  A.~G. Stefani, J.~B. Huber, C. Jardin and H. Sticht, ``Towards 
  confidence intervals for the mutual information between two binary
  random variables,'' In \emph{Proc. Workshop Computational Systems
  Biology (WCSB 2012)}, Ulm, Germany, Jun. 4--6, 2012.
\bibitem{Weissman2003}
  T. Weissman, E. Ordentlich, G. Seroussi, S. Verd\'{u} and M.J.
  Weiberger, ``Inequalities for the L1 Deviation of the Empirical
  Distribution,'' Tech. Rept., HP Laboratories Palo Alto, HPL-2003-97
  (R.1), Jun. 2003.
\bibitem{Othersen2012}
  O.~G. Othersen, A.~G. Stefani, J.~B. Huber and H. Sticht, ``Application
  of Information Theory to Feature Selection in Protein Docking,'' \emph{
  J Mol Model.}, vol.~18, no. 4, pp. 1285--1297, Jul. 2012.
\bibitem{Stefani2013}
  A.~G. Stefani, J.~B. Huber, C. Jardin and H. Sticht, ``A tight lower
  bound on the mutual information of a binary and an arbitrary finite
  random variable in dependence of the variational distance,'' available
  at http://arxiv.org/abs/1301.5937.
\end{thebibliography}

\end{document}